\documentclass{jps-cp}
\usepackage{txfonts} 
\usepackage{color}

\title{Search for the field-induced magnetic instability around the upper critical field of superconductivity in $H \parallel c$ in CeCoIn$_{5}$}

\author{Takanori  \textsc{Taniguchi}$^{1}$, Shunsaku \textsc{Kitagawa}$^{1}$, Masahiro \textsc{Manago}$^{1}$, Genki \textsc{Nakamine}$^{1}$, Kenji \textsc{Ishida}$^{1}$ and Hiroaki \textsc{Shishido}$^{2}$}

\inst{$^{1}$Department of Physics, Kyoto University, Kyoto 606-8502, Japan \\
$^{2}$Department of Physics and Electronics, Osaka Prefecture University, Osaka 980-8577, Japan}

\email{taka.taniguchi@imr.tohoku.ac.jp}

\recdate{February 21, 2019}

\abst{We present nuclear spin-lattice relaxation rate ($1/T_{\rm 1}$) at the Co site and ac-susceptibility results in the normal and superconducting (SC) states of CeCoIn$_{\rm 5}$ for $H \parallel c$ near the SC upper critical field $H_{\rm c2}$ above 0.1 K. 
At 4.2 T, $1/T_{\rm 1}$ rapidly decreases below the SC transition temperature, consistent with the previous reports. 
Although the field dependence of $1/T_{\rm 1}T$ at 0.1 K shows a peak at 5.2 T above $H_{\rm c2}$, the temperature dependence of $1/T_{\rm 1}T$ at 5.2 T is independent of temperature below 0.2 K, showing a Fermi-liquid behavior. 
In addition, we found no NMR-spectrum broadening by the appearance of internal fields around $H_{\rm c2}$ at 0.1 K.
We could not detect any field-induced magnetic instability around $H_{\rm c2}$ down to 0.1 K although the remarkable non-Fermi-liquid behavior towards $H_{\rm c2}$ was observed in various physical quantities. }

\kword{field-induced criticality, NMR, ac susceptibility, heavy fermion}

\begin{document}
\maketitle

\section{Introduction}

Heavy fermion systems have many examples of a quantum critical point (QCP), where magnetic transition temperature is suppressed to zero by applied a pressure or magnetic field, and unconventional superconductivity has been discovered in many cases. 
This fact provides strong evidence for magnetically mediated Cooper pairing in these superconductors\cite{Stewart_1984,Stewart_2011,Gegenwart_2008}.
In particular, Ce$T$In$_{\rm 5}$ ($T=$ Co, Ir, Rh) series, called 1-1-5 family, have attracted much attention for QCP studies because of an emergence of various phenomena originating from 4$f$ electrons such as antiferromagnetic (AFM) order, heavy fermion (FL) state, and superconducting (SC) state under the pressure or magnetic field \cite{Park_2006,Sakai_2011,Thompson_2012}. 

In this paper, we focus on CeCoIn$_{\rm 5}$. 
This compound is located in the vicinity of the AFM phase and shows superconductivity at a transition temperature $T_{\rm c} = 2.3$ K at zero-field and ambient pressure \cite{Petrovic_2001}. 
The Knight shift decreased rapidly on entering the SC phase for $ H \parallel c$, indicating a singlet Cooper pair forming in CeCoIn$_5$\cite{Curro_2001, Kohori_2001}. 
Nuclear spin-lattice relaxation rate ($1/T_{\rm 1}$) was proportional to $\thicksim T^{\rm 3}$ far below $T_{\rm c}$, suggesting that CeCoIn$_5$ is a line-nodal superconductor \cite{Kohori_2001}. 
In addition, the $d_{\rm x^{2}-y^{2}}$ symmetry was suggested from the several measurements such as the field-angle resolved thermal-conductivity\cite{Izawa_2001} and heat-capacity \cite{An_2010} measurements and the spectroscopic imaging scanning tunneling microscopy \cite{Allan_2013,Zhou_2013} measurement. 

For $H \parallel ab$, coexistence of Flude-Ferrell-Larkin-Ovchinnikov (FFLO) state and incommensurate spin-density-wave order (called Q-phase) was observed near $H_{\rm c2}$ \cite{Young_2007, Kenzelmann_2008}. 
The NMR spectrum at the In(2) site abruptly shifts at $T_c$ with the first order character determined with specific-heat and magnetization measurements near $H_{c2}$, and splits in the Q-phase inside the SC state\cite{Young_2007,Bianchi_2003,Murphy_2002,Tayama_2002}.
It is noted that the magnetic Q-phase can exist inside the SC state.
On the other hand, the presence of pure FFLO state in $H \parallel c$ was suggested from the various measurements near $H_{\rm c2}$ \cite{Tayama_2002,Kumagai_2009}, but the crucial evidence of the FFLO state in $H \parallel c$ has not been obtained. 
In addition, the presence of the field-induced magnetic instability similar to the Q phase is another interesting issue for $H \parallel c$ near $H_{\rm c2}$, since a remarkable non Fermi liquid behavior towards $H_{\rm c2}$ was observed in various measurements.
Sakai $et$ $al$. reported that $1/T_{\rm 1}T$ measured at 5 T increased down to 0.2 K with decreasing temperature due to the development of the quasi-2D spin fluctuations \cite{Sakai_2011}, but $1/T_{\rm 1}T$ below 0.2 K was constant (Korringa-law), indicating the FL state. 
This was consistent with the dHvA \cite{Settai_2001}, resistibility \cite{Ronning_2005}, and thermal conductivity measurements \cite{Izawa_2007}. 
In the NMR study by Sakai {\it et al.}, the presence of the maximum of $1/T_{\rm 1}T$ near $H_{\rm c2}$ was shown and suggested the possibility of the coincidence with the field-induced magnetic critical point and $H_{\rm c2}$ in $H \parallel c$, but the detailed search of the magnetic phase has not been performed\cite{Sakai_2011}. 
In this paper, we report the detailed temperature and field dependence of $^{59}$Co-NMR and ac susceptibility results for $H \parallel c$ near $H_{\rm c2}$.
No superconductivity was observed from the ac susceptibility measurement above 5.0 T and 0.2 K, but $1/T_{\rm 1}T$ at 0.1 K shows a maximum at 5.2 T above $H_{c2}$. 
However, $1/T_{\rm 1}T$ measured at 5.2 T stayed constant below 0.2 K, indicative of the absence of field-induced magnetic instability around $H_{\rm c2}$. 

\section{Experimental Procedures}
Single crystals of CeCoIn$_{\rm 5}$ were synthesized by the In-flux method\cite{Shishido_2002}. 
We selected the thin-plate single crystal (1.5 mm $\times$ 2.0 mm $\times$ 0.2 mm) with the $c$ axis normal to the plate, and performed the $^{59}$Co-NMR measurement on the single crystal.
The $^{59}$Co-NMR spectra were obtained by summing the Fourier transform spectra from the spin-echo signal obtained at equally spaced rf frequencies at a fixed magnetic field. 
Since $^{59}$Co nuclear spin is $7/2$ ($I = 7/2$), the $^{59}$Co-NMR spectrum consists of seven peaks. 
The sharpness of the resonance line indicates the high quality of the crystal. 
$1/T_{\rm 1}$ at the Co site was obtained from a central line arising from the $\left|-1/2\right\rangle \Longleftrightarrow\left|+1/2\right\rangle $ transition. 
The crystal alignment with respect to the magnetic field was precisely performed with a single-axis rotator in the horizontal field generated by a split-magnet, and the misalignment is within $0.5^{\circ}$.

\section{Results}
\begin{figure}[tbh]
\includegraphics[scale=0.95]{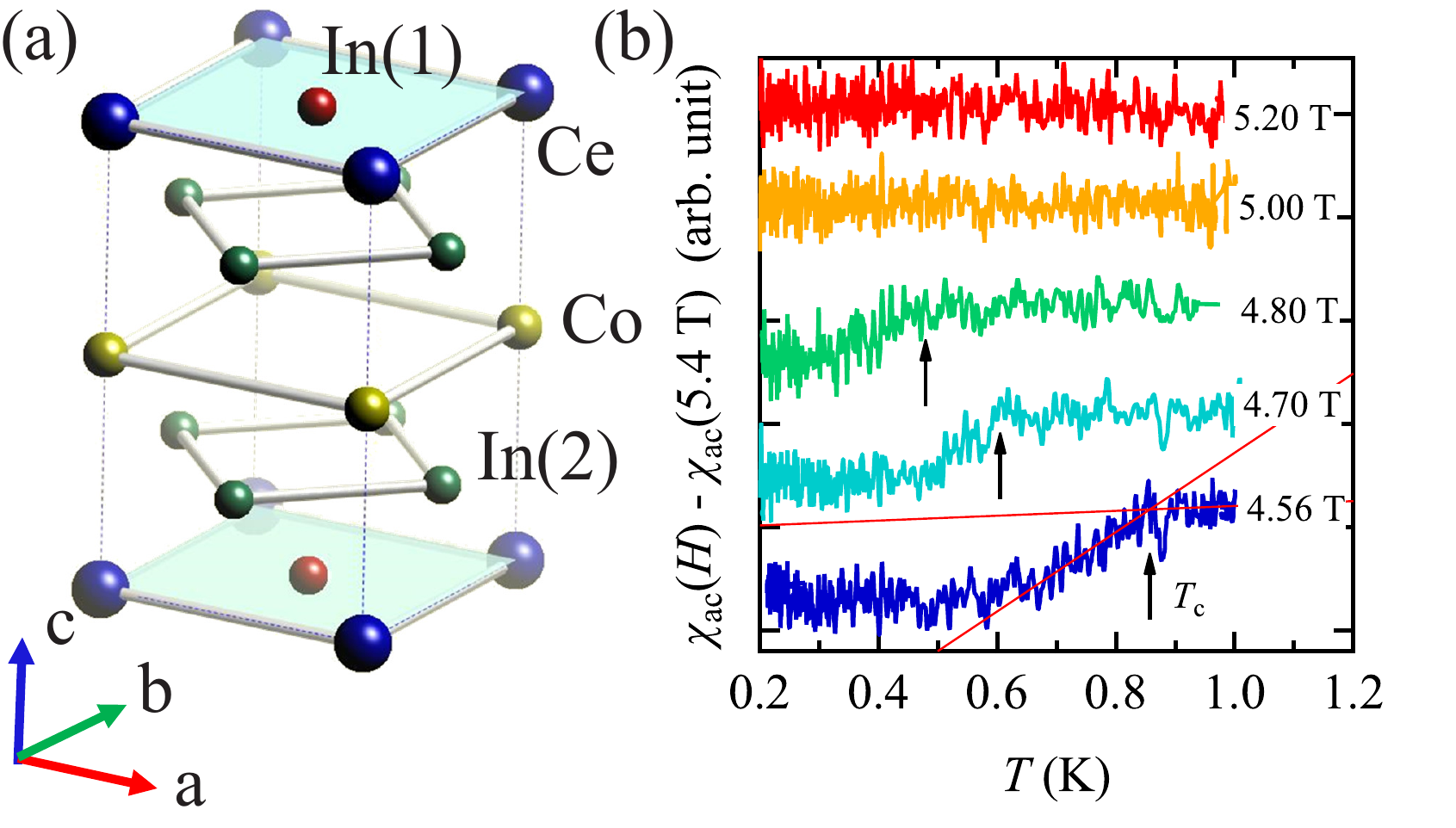}
\caption{(a) The crystal structure of CeCoIn$_{5}$. The blue semitransparent sheets present the Ce and In(1) plane. (b) The temperature dependence of the $\chi_{\textrm{ac}}$ near $H_{\rm c2}$. For the sake of clarity, the plots in (b) are vertically shifted consecutively. $T_{\rm c}$ is determined by liner fitting of the diamagnetic signals around the turning point, like the red lines at 4.56 T, and is indicated the black arrows.}
\label{f1}
\end{figure}

Figure 1(a) shows the crystal structure of CeCoIn$_{\rm 5}$. 
The crystal structure of CeCoIn$_5$ is the tetragonal HoCoGa$_{\rm 5}$ type structure and the space group $P$4/$mmm$\cite{Petrovic_2001}. 
The Ce and Co sites, occupying the 1a and 1b Wyckoff positions with the both point symmetries 4/$mmm$, are at the same fourfold axis along the $c$ axis. 
There are two crystallographycally inequivalent In sites, In(1) and In(2), occupying the 1c and 4i Wyckoff positions with the point symmetries 4/$mmm$ and  2$mm$., respectively.

Figure 1(b) shows the temperature dependence of ac susceptibility $\chi_{\textrm{ac}}$ for various magnetic fields parallel to the $c$ axis. 
Below 4.8 T, $\chi_{\textrm{ac}}$ decreases below $T_c(H)$, caused by the diamagnetic shielding. 
On the other hand, above 5.0 T, a clear anomaly is not observed down to 0.2 K, which is consistent with the previous reports ($\mu_0 H_{\rm c2}$ = 5.0 T) \cite{Petrovic_2001}.

\begin{figure}[tbh]
\includegraphics[scale=0.75]{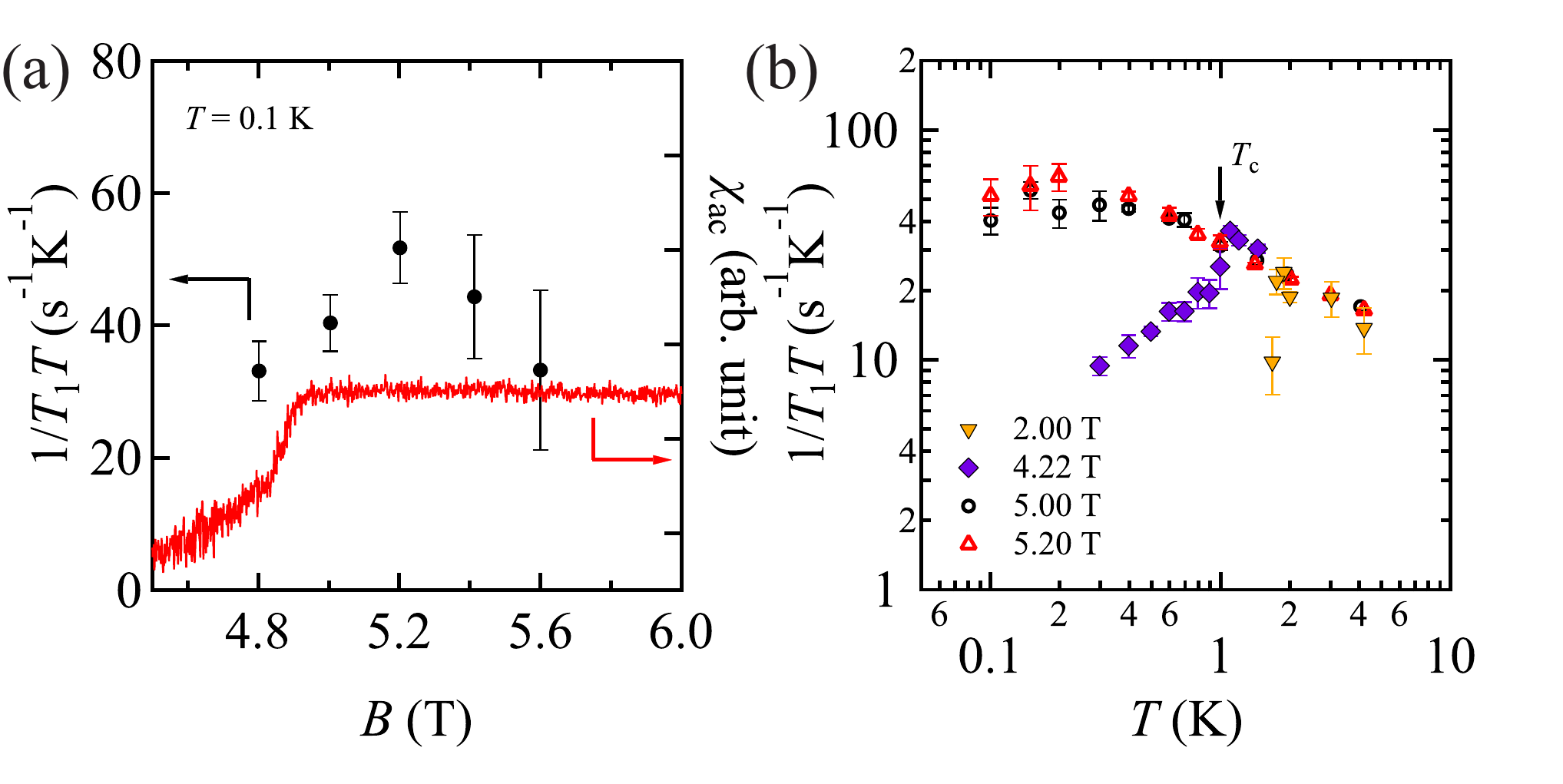}
\caption{(a) The field dependence of $1/T_{\rm 1}T$ and ac magnetic susceptibility at 0.1 K. (b) The temperature dependence of $1/T_{\rm 1}T$ at several fields. The black arrow shows the $T_{\rm c}$s at 4.22 T determined by the ac susceptibility measurement.}
\label{f2}
\end{figure}

Figure 2(a) shows the field dependence of $1/T_{\rm 1}T$ and $\chi_{\textrm{ac}}$ at 0.1 K near $H_{\rm c2}$ for $H \parallel c$. 
At 4.8 T, the $^{59}$Co-NMR spectrum was splitting below $T_{\rm c}$ due to the first order transition, and we follow the normal-state component. 
$1/T_{\rm 1}T$ showed a peak at 5.2 T, suggesting that the field-induced peak of $1/T_1T$ is above $H_{\rm c2}$. 
Figure 2(b) shows the temperature dependence of $1/T_{\rm 1}T$ in various magnetic fields parallel to $c$ axis. 
Above $T_{\rm c}$, $1/T_{\rm 1}T$ increases with decreasing temperature due to the development of AFM fluctuations.  
In smaller fields than $H_{\rm c2}$ (2.0 and 4.2 T), $1/T_{\rm 1}T$ abruptly decreases without a coherent peak just below $T_c$, supporting the $d$-wave superconductivity. 
At 5.2 T, $1/T_{\rm 1}T$ is independent of temperature below 0.2 K, showing the Korringa behavior even at the critical field where $1/T_1T$ shows a maximum. 
This indicates that a field-induced magnetic instability is absent in CeCoIn$_5$ near $H_{\rm c2}$ in $H \parallel c$.

\begin{figure}[tbh]
\includegraphics[scale=0.75]{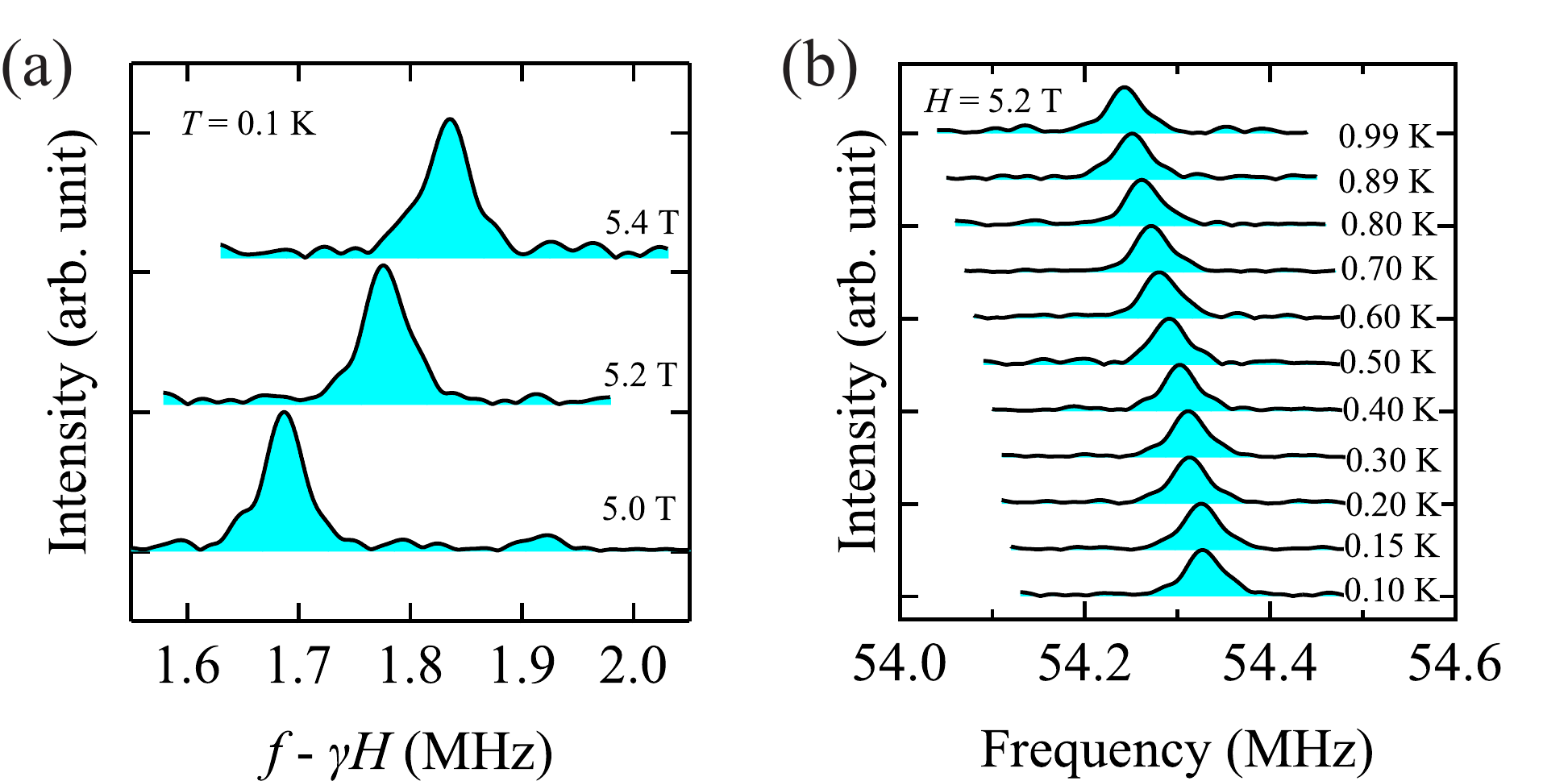}
\caption{(a) $^{59}$Co-NMR spectra measured at 0.1 K in 5.0, 5.2 and 5.4 T. (b) The temperature dependence of $^{59}$Co-NMR spectra at 5.2 T. Panels (a) and (b) show the Co-NMR center line arising from $\left|-1/2\right\rangle \Longleftrightarrow\left|+1/2\right\rangle $ transition.}
\label{f3}
\end{figure}

Figure 3 (a) shows the $^{59}$Co-NMR spectra measured at 0.1 K in 5.0, 5.2, and 5.4 T near $H_{\rm c2}$. 
The shape of lines are insensitive to the field, and any linewidth broadening of the $^{59}$Co-NMR spectra was not observed at 5.2 T, as shown the Fig. 3(b), also indicative of the absence of due to the appearance of static internal field a field-induced magnetic instability around $H_{\rm c2}$ as well as the result of $1/T_{\rm 1}T$.

Many anomalous states due to the development of spin-fluctuations, such as a non-Fermi liquid behavior, have been observed near QCPs in various heavy-fermion compounds. 
A typical example is YbRh$_{\rm 2}$Si$_{\rm 2}$, in which $1/T_{\rm 1}T$ of $^{29}$Si in YbRh$_2$Si$_2$ continues to increase at the field-induced critical point with decreasing temperature\cite{Ishida_2002}. 
However, our result shows that $1/T_1T$ is constant below 0.2 K, indicating that the FL state holds even at the critical field where $1/T_1T$ shows the maximum. 
The $1/T_1T$ behavior observed in the critical region of CeCoIn$_5$ is reminiscent of the criticality observed in CeRu$_2$Si$_2$ showing the meta-magnetic criticality. 
Sakakibara {\it et al.} reported that the FL state holds at low temperatures even at the critical region, and suggested that the origin of this criticality is related with the change of the $f$ electron character from itinerant to localized \cite{Sakakibara_1995}.
$1/T_{\rm 1}T$ of CeRu$_{2}$Si$_{2}$ at the critical field increased down to $T_{\rm M}$ and stayed constant below $T_{\rm M}$\cite{Ishida_1998}, as if CeRu$_2$Si$_2$ possesses a finite crossover temperature at $T_{\rm M}$\cite{Sakakibara_1995}.
We suggest that the critical behavior observed in CeCoIn$_5$ near $H_{\rm c2}$ might originate from the field continuous change of the Fermi-surfaces, in which the large Fermi-surfaces still remain above $H_{\rm c2}$ as shown in \cite{Settai_2001}. 
The $1/T_1T$ behaviors against $H$ and $T$ show the crossover from the non-FL to FL occurring at around 5.2 T below 0.2 K, in good agreement with the recent thermal expansion measurement\cite{Zaum_2011}, but inconsistent with the thermal conductivity results showing the non-FL behaviors above $H_{\rm c2}$\cite{Izawa_2007}.

Recently, the field-induced phase transition was reported at around 8 T and 15 mK from the dHvA measurement, but the order parameter is unknown yet \cite{Shishido_2018}. 
From the present $^{59}$Co-NMR measurement, it seems that the field-induced phase transition is not ascribed to magnetic in origin but would originate from the Fermi-surface instability probed by the transport measurement, since AFM magnetic fluctuations are suppressed at the higher fields\cite{Sakai_2011}. 
However it is noted that the phase-transition was observed up to 9.5 T and seems to exist above 10 T in $H \parallel c$, which is anticipated to the relation with the magnetic Q phase observed in $H \perp c$.    
To understand the origin of the field-induced transition, further measurements, particularly angle dependence of the NMR measurements in the $ac$ plane down to 10 mK would give a clue to clarify the origin.


\section{Conclusion}
We have performed the single crystal $^{59}$Co-NMR measurements for $H \parallel c$ near $H_{\rm c2}$ in CeCoIn$_{\rm 5}$ to investigate whether the field induced magnetic phase like the Q phase is observed or not. 
Although low temperature $1/T_{\rm 1}T$ shows a peak at 5.2 T above $H_{\rm c2}$, $1/T_{\rm 1}T$ at 5.2 T is almost constant below 0.2 K, indicative of an absence of field-induced magnetic instability down to 0.1 K.

\section*{Acknowledgment}
We would like to thank Y. Maeno and S. Yonezawa for stimulating discussions. This work was financially supported by JSPS/MEXT Grants-in-Aids for Scientific Research (KAKENHI) Grant Numbers, JP15H05745, JP17K14339, JP19K03751, JP19K14657, JP19H04696, JP19K23417, Grant Numbers JP15H05882, JP15H05884 (J-Physics). T. Taniguchi and M. Manago were supported by the JSPS Research Fellowship (JP17J08806 and JP17J05509 , respectively),

\end{document}